\documentclass[jou]{interact}\usepackage[]{graphicx}\usepackage[]{color}
\makeatletter
\def\maxwidth{ %
  \ifdim\Gin@nat@width>\linewidth
    \linewidth
  \else
    \Gin@nat@width
  \fi
}
\makeatother

\definecolor{fgcolor}{rgb}{0.345, 0.345, 0.345}

\usepackage{framed}
\makeatletter
 {\par\unskip\endMakeFramed%
 \at@end@of@kframe}
\makeatother

\definecolor{shadecolor}{rgb}{.97, .97, .97}
\definecolor{messagecolor}{rgb}{0, 0, 0}
\definecolor{warningcolor}{rgb}{1, 0, 1}
\definecolor{errorcolor}{rgb}{1, 0, 0}

\usepackage{alltt}

\usepackage{dsfont}
\usepackage{xcolor}
\usepackage{tcolorbox}
\usepackage{cancel}

\usepackage[english]{babel}
\usepackage{inputenc}

\usepackage{multicol, amsthm}

\usepackage{parskip}
\usepackage{scalerel,stackengine}
\stackMath
\newcommand\reallywidehat[1]{%
\savestack{\tmpbox}{\stretchto{%
  \scaleto{%
    \scalerel*[\widthof{\ensuremath{#1}}]{\kern-.6pt\bigwedge\kern-.6pt}%
    {\rule[-\textheight/2]{1ex}{\textheight}}
  }{\textheight}%
}{0.5ex}}%
\stackon[1pt]{#1}{\tmpbox}%
}

\usepackage{booktabs,array}

\newcount\rowc

\makeatletter
\def\ttabular{%
\hbox\bgroup
\let\\\cr
\def\rulea{\ifnum\rowc=\@ne \hrule height 1.3pt \fi}
\def\ruleb{
\ifnum\rowc=1\hrule height 1.3pt \else
\ifnum\rowc=6\hrule height \heavyrulewidth 
   \else \hrule height \lightrulewidth\fi\fi}
\valign\bgroup
\global\rowc\@ne
\rulea
\hbox to 10em{\strut \hfill##\hfill}%
\ruleb
&&%
\global\advance\rowc\@ne
\hbox to 10em{\strut\hfill##\hfill}%
\ruleb
\cr}
\def\endttabular{%
\crcr\egroup\egroup}

\usepackage{epstopdf}
\usepackage[caption=false]{subfig}

\usepackage[doublespacing]{setspace}
\usepackage{inputenc}

\theoremstyle{plain}

\theoremstyle{definition}

\theoremstyle{remark}

\usepackage{natbib}
\IfFileExists{upquote.sty}{\usepackage{upquote}}{}
\begin{document}
\bibliographystyle{apalike}

\articletype{Original Research Paper}

\title{A non-inferiority test for $R^{2}$ with random regressors}

\author{
\name{Harlan Campbell\thanks{CONTACT: Harlan Campbell. Email: harlan.campbell@stat.ubc.ca}}
\affil{University of British Columbia Department of Statistics
Vancouver, BC, Canada, V6T 1Z2}
}

\maketitle

\begin{abstract}
Determining the lack of  association between an outcome variable and a number of different explanatory variables is frequently necessary in order to disregard a proposed model.  This paper proposes a non-inferiority  test for the coefficient of determination (or squared multiple correlation coefficient), $R^{2}$, in a linear regression analysis with random predictors.  The test is derived from inverting a one-sided confidence interval based on a scaled central F distribution. 
\end{abstract}
 \begin{keywords}
equivalence testing, non-inferiority testing, linear regression, standardized effect sizes
\end{keywords}

\newpage

\section{Introduction}

 The coefficient of determination (or squared multiple correlation coefficient), $R^{2}$, is a well-known and well-used statistic for linear regression analysis.  $R^{2}$ summarizes the ``proportion of variance explained'' by the predictors in the linear model and is equal to the square of the Pearson correlation coefficient between the observed and predicted outcomes \citep{nagelkerke1991note, zou2003correlation}.  Despite the $R^{2}$ statistic's ubiquitous use, its corresponding population parameter, which we will denote as $P^{2}$,  as in \cite{cramer1987mean}, is rarely discussed.  $P^2$ is  sometimes known as the ``parent multiple correlation coefficient'' \citep{barten1962note} or the ``population proportion of variance accounted for'' \citep{kelley2007confidence}; see \cite{cramer1987mean} for details.

 \citet{campbelllakens2019} introduced a non-inferiority test (a one-sided equivalence test) for $P^{2}$ in order to test the hypotheses:
 
 $H_{0}: 1 > P^{2} \ge \Delta$,\\
 \indent $H_{1}: 0 \le P^{2} < \Delta$;

 \noindent where $[0, \Delta]$ is the non-inferiority margin representing a range of effect sizes of negligible magnitude.  The test is useful for determining whether one can reject the hypothesis that the total proportion of variance in the outcome, $Y$, attributable to the set of covariates, $X$, is greater than or equal to $\Delta$.  Or phrased somewhat differently, the test asks whether we ``can disregard the whole model''?   \citep{campbelllakens2019}.

 \citet{campbelllakens2019} compared their frequentist non-inferiority test with a Bayesian approach based on Bayes Factors and also provided a version of the test for the $\eta^{2}$ parameter in a fixed effects (or ``between subjects") analysis of variance (ANOVA). However, the non-inferiority test put forward only applied to cases with fixed regressors.  The sampling distribution of $R^{2}$ can be quite different when regressor variables are random; see  \cite{gatsonis1989multiple}. 
 
Indeed, depending on whether regressors are fixed or random, certain inference procedures for $P^{2}$ will be different.  Random regressors are more common in observational studies, whereas fixed regressors are more common in experimental studies where the regressors are randomized by experimenters or otherwise fixed by some study intervention.  For a standard null hypothesis significance test (i.e.,  $H_{0}: P^{2} = 0$), the same central $F$-distributed statistic can be used for random regressors and fixed regressors.  This is due to the fact that when the null hypothesis is true, the sampling distribution of $R^{2}$ is identical for both cases.  However, when $P^{2} \ne 0$, the sampling distribution of $R^{2}$ does indeed depend on whether the regressors are fixed or random.  
 
 In this short article, we propose a $P^{2}$ non-inferiority test for situations with random regressors.  In the social sciences and many other fields of study, the assumption of fixed regressors is often violated and therefore it is important to consider for this possibility \citep{bentler1983covariance}.  In Section 2, we describe the proposed test and in Section 3 we conduct a small simulation study to examine the test's operating characteristics. 
 
 \section{A non-inferiority test for random regressors}

Let $N$ be the number of observations and $K$ be the number of covariates in a standard multivariable linear regression analysis.  Let $Y_{i}$ be the outcome variable for the $i$-th subject and $X_{i}$ be the vector of covariates, $(X_{1},..., X_{K}$, for the $i$-th subject.  Then the matrix $X$ is a $N$ by $K$ design matrix and the linear regression model can be summarized by:

\begin{equation} Y_{i} = \beta_{0} + X_{i}\beta + \epsilon    _{i}, \quad \textrm{where:} \quad  \epsilon_{i}\sim  \; \mathcal{N}(0, \sigma^{2}), \quad \quad \forall \; i=1,...,N;
\label{regress}
\end{equation}

 \noindent where $\beta  = (\beta_{1} , ... , \beta_{K} )^{'}$ is the column-vector of regression coefficients and $\sigma^{2}$ is the residual variance.  
 
As mentioned in the Introduction, we are specifically interested in the scenario of ``random regressors,'' in which the covariates, $X_{1},...,X_{K}$, are assumed to be stochastic rather than fixed.  In practice, the assumption of ``fixed regressors'' would be more appropriate for a randomized trial, whereas the assumption of ``random regressors'' would be more appropriate for an observational study.  We require that the rows of X be independent of each other and independent of $\epsilon_{i}$.

  A non-inferiority test $p$-value can be obtained by inverting a one-sided confidence interval.  However, constructing a confidence interval for $P^{2}$ with random regressors is not at all obvious.  Several procedures have been proposed in the literature. These include Wald-type confidence intervals and bootstrap-based confidence intervals \citep{tan2012confidence}.  However, neither of these approaches have particularly good finite sample properties; see \citet{algina1999comparison}.
 
 \citet{helland1987interpretation} proposes obtaining a confidence interval for $P^{2}$ by relying on a scaled central $F$ approximation of $\tilde{P}^{2} = P^{2}/(1-P^{2})$, and provides a simple iterative procedure that provides ``surprisingly good'' \citep{helland1987interpretation} accuracy.  \cite{tan2012confidence} agrees. After reviewing a number of alternative methods, \cite{tan2012confidence} concludes that ``the scaled central $F$ approximation [method] seems to be a simple and good procedure to construct an asymptotic confidence interval.''  We will therefore use this proposed confidence interval, inverted, for our non-inferiority test.   Note that the scaled central $F$ approximation method is based on the assumption that the covariate matrix $X$ has a multivariate normal distribution.

 For a given value for $\alpha$ (e.g.,  $\alpha=0.10$), and taking for an initial value, $R^{2*}_{\alpha} = R^{2}$, we can obtain a one-sided confidence interval for $P^{2}$ (e.g., a one-sided upper 90\% CI) by iterating between calculating $v$ and $R^{2*}_{\alpha}$ until convergence, where:
 
 \begin{equation}
     v = \frac{\Big((N-K-1)R^{2*}_{\alpha} + K \Big)^{2} }{N - 1 - (N-K-1)(1-R^{2*}_{\alpha})^{2}}
 \end{equation}
 
 and 
 
 \begin{equation}
     R^{2*}_{\alpha} = \frac{(N-K-1)R^{2} - (1-R^{2})KF_{\alpha, v, (N-K-1)}}{(N-K-1)\Big[R^{2} + (1-R^{2})F_{\alpha, v, (N-K-1)}\Big]},
 \end{equation}
 
  \noindent where $F_{\alpha, v, (N-K-1)}$ is the $\alpha$\% percentile of the central $F$ distribution with $v$ and $n-p-1$ degrees of freedom.  
  
 We then calculate the upper ($1-\alpha$)\% confidence interval,  $\textrm{CI}(P^{2})_{(1-\alpha)\%}$, as follows:
 
 \begin{equation}
 \textrm{CI}(P^{2})_{(1-\alpha)\%} = \Big[0, \frac{(N-K-1)R^{2} - (1-R^{2})KF_{\alpha, v, (N-K-1)}} { (N-K-1)( R^{2} + (1-R^{2})F_{\alpha, v, (N-K-1)})} \Big] .
 \label{eq:CI}
  \end{equation}

Note that in the R package ``MBESS'' \citep{kelley2007confidence}, the function ci.R2 can be used to calculate a one-sided confidence interval for $P^{2}$ with random regressors.  This calculation is based on the scaled \emph{non}-central $F$ approximation of \cite{lee1971some} and, in our experience, will provide a very similar result.  Note that there is also SAS code and SPSS code made available from \citet{zou2007toward} for the  calculation of confidence intervals based on the scaled \emph{non}-central F approximation. 

 
 
 In order to obtain a $p$-value for a non-inferiority test ($H_{0}: 1 > P^{2} \ge \Delta$), we must invert the upper one-sided confidence interval.  We proceed as follows.  First, we calculate the following $F$-statistic:

 \begin{equation}
     F_{\Delta} = \frac{(N-K-1)R^{2}(\Delta-1)}{(R^{2}-1)(\Delta(N-K-1) + K )}
 \end{equation}
 
 We then iterate between calculating $v$ and $R^{2*}$ until convergence:

\begin{equation}
     v = \frac{\Big((N-K-1)R^{2*} + K \Big)^{2} }{N - 1 - (N-K-1)(1-R^{2*})^{2}}
 \end{equation}
 
 and 
 
 \begin{equation}
     R^{2*} = \frac{(N-K-1)R^{2} - (1-R^{2})KF_{\Delta}}{(N-K-1)\Big[R^{2} + (1-R^{2})F_{\Delta}\Big]}.
 \end{equation}

The $p$-value for the non-inferiority test can then be calculated as: 

\begin{equation}
p-\textrm{value} = p_{f}\left(F; v, N-K-1 \right),
\label{eq:pval}
\end{equation}

 \noindent where $p_{f}(\cdot \quad ; df_{1}, df_{2})$ is the cdf of the central $F$-distribution with $df_{1}$ and $df_{2}$ degrees of freedom.  It is important to remember that the above test makes the assumption that the residuals and the regressors are independent of one another and that both are normally distributed.

 \section{Simulation Study}

We conducted a simple simulation study in order to better understand the operating characteristics of the non-inferiority test and to confirm that the test has correct type 1 error rates.  We followed a very similar design for the simulation study as \cite{campbelllakens2019}.  We simulated data for each of \textcolor{black}{thirty} scenarios, one for each combination of the following parameters:
 
\begin{itemize}
     \item one of three variances: $\sigma^{2}=0.4$, $\sigma^{2}=0.5$, or $\sigma^{2}=1.0$;

    \item one of five sample sizes: $N=60$, $N=180$, $N=540$, $N=1,000$, or, $N=8,000$;
    
    \item one of two values for $K=2$, or $K=4$; with $\beta=(0.11, -0.15)$ or $\beta=(0.11,  0.10, -0.05, -0.10)$, ($\beta_{0}=0$ for all scenarios). The covariates values are sampled from a multivariate normal distribution.  For $K=2$, we have:
\begin{equation*}
 X_{i}
\sim \mathcal{MVN} \begin{pmatrix}
(0,0),
\begin{pmatrix}
1.00, \quad 0.05 \\
0.05, \quad 1.00 \\
 \end{pmatrix}
\end{pmatrix}, \forall i = 1,..,N.
\end{equation*}

\noindent For $K=4$, we have:

\begin{equation*}
 X_{i}
\sim \mathcal{MVN} \begin{pmatrix}
(0,0,0,0),
\begin{pmatrix}
1.00, \quad 0.05 , \quad 0.05 , \quad 0.05 \\
0.05, \quad 1.00 , \quad 0.05 , \quad 0.05 \\
0.05, \quad 0.05 , \quad 1.00, \quad 0.05 \\
 0.05,\quad 0.05, \quad 0.05 , \quad 1.00 \\
 \end{pmatrix}
\end{pmatrix}, \forall i = 1,..,N.
\end{equation*}

\end{itemize}

\noindent For each single simulated dataset, we sampled a new $X$ matrix from the chosen multivariate normal distribution.  Depending on the particular values of $K$ and $\sigma^{2}$, the true coefficient of \textcolor{black}{determination} for these data is either $P^{2}=0.034$, $P^{2}=0.065$, or $P^{2}=0.080$.    Parameters for the simulation study were chosen so as to obtain three unique values for $P^{2}$ approximately evenly spaced between 0 and 0.10.

\textcolor{black}{ For each of the thirty configurations, we simulated 50,000 unique datasets and calculated a non-inferiority $p$-value with each of 19 different values of $\Delta$ (ranging from 0.01 to 0.10).  We then calculated the proportion of these $p$-values less than $\alpha=0.05$.  }



\begin{figure}
    \centering
    \includegraphics{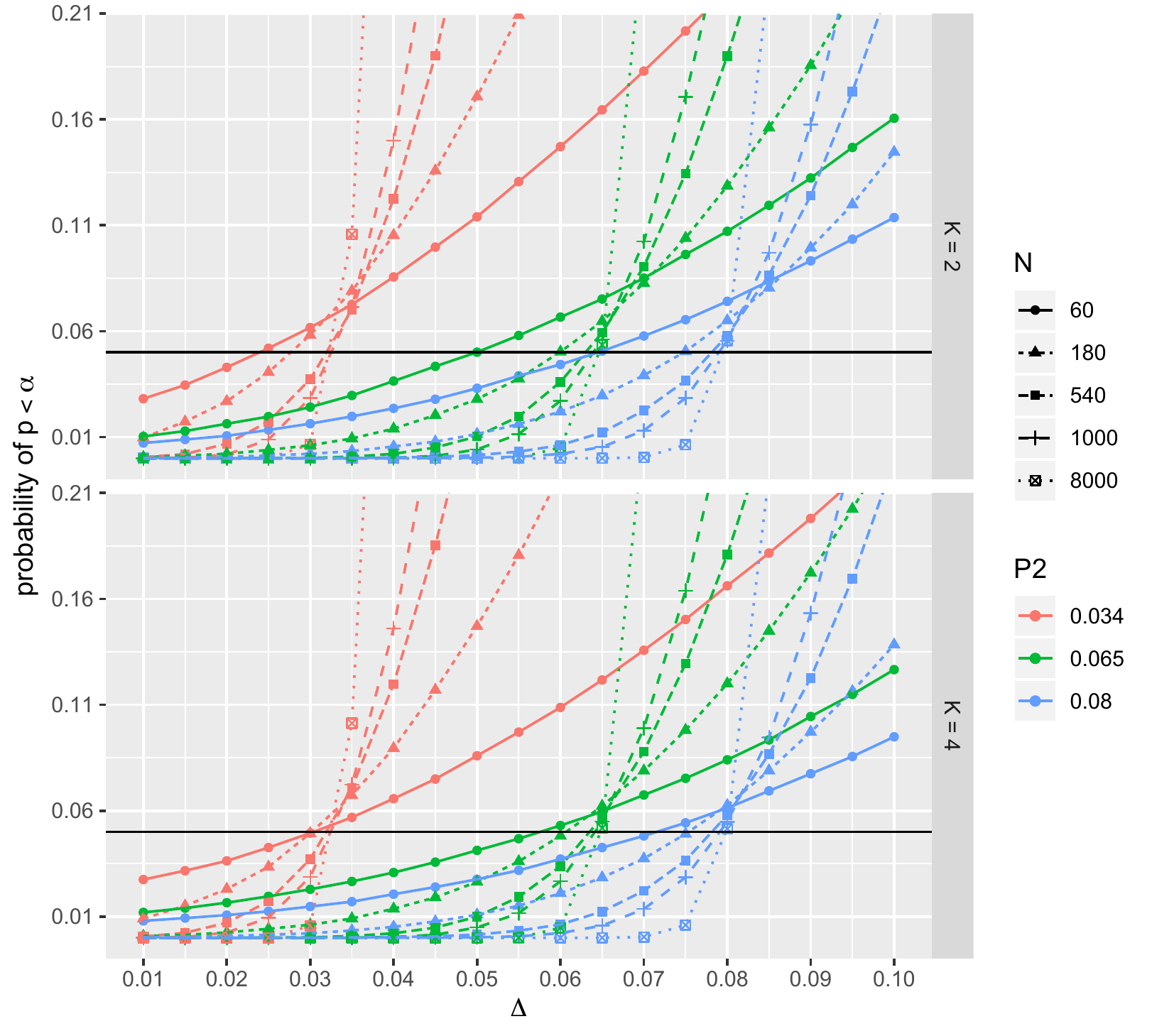}
 \caption{Simulation study results.  Upper panel shows results for $K=2$; lower panel shows results for $K=4$.  Both plots are presented with a restricted vertical-axis to better show the type 1 error rates.  The solid horizontal black line indicates the desired type 1 error of $\alpha=0.05$.   For each of thirty configurations, we simulated 50,000 unique datasets and calculated a non-inferiority $p$-value with each of 19 different values of $\Delta$ (ranging from 0.01 to 0.10).}
    \label{fig:trunc}
\end{figure}

\begin{figure}
    \centering
    \includegraphics{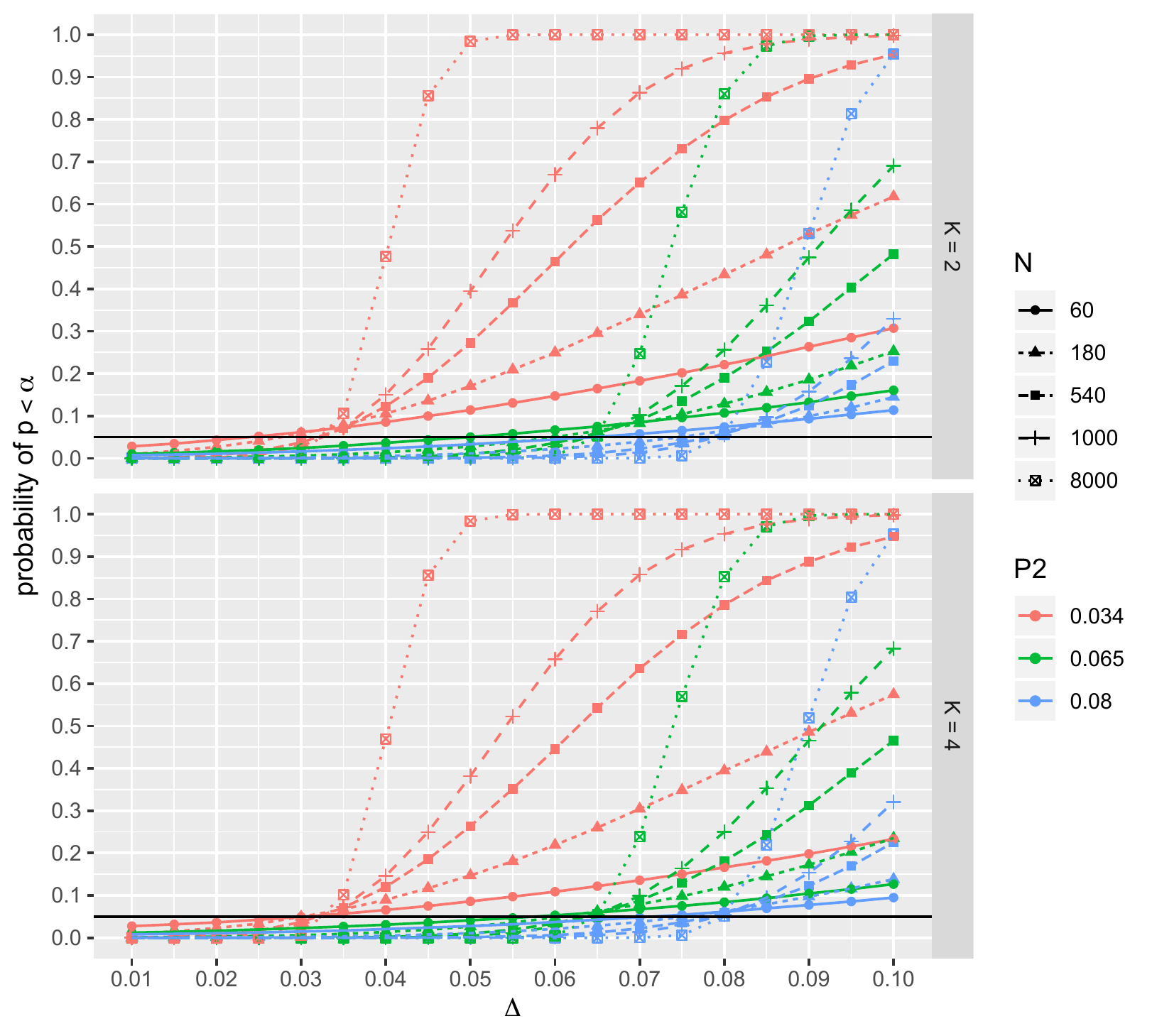}
    \caption{Simulation study, complete results.  Upper panel shows results for $K=2$; Lower panel shows results for $K=4$.  The solid horizontal black line indicates the desired type 1 error of $\alpha=0.05$.  For each of thirty configurations, we simulated 50,000 unique datasets and calculated a non-inferiority $p$-value with each of 19 different values of $\Delta$ (ranging from 0.01 to 0.10).}
    \label{fig:full}
\end{figure}

Figures \ref{fig:trunc} and \ref{fig:full} plot the results.  Note that Figure \ref{fig:trunc} is on restricted vertical axis to better show the type 1 error rates.  We see that when the non-inferiority bound $\Delta$ equals the true effect size (i.e., 0.034, 0.065, or 0.080), the type 1 error rate is exactly 0.05, as it should be, for all moderately large values of $N$.  This situation represents the boundary of the null hypothesis, i.e. $H_{0}: \Delta \le P^{2}$.  When $N$ is smaller (i.e., when $N=60$ or $N=180$), the type 1 error is slightly larger than the desired rate of $\alpha=0.05$ when $\Delta$ equals the true effect size.  

As the equivalence bound increases beyond the true effect size (i.e., $\Delta > P^{2}$), the alternative hypothesis is then true and it becomes possible to correctly reject the null. \textcolor{black}{
As expected, the power of the test increases with larger values of $\Delta$, larger values of $N$, and smaller values of $K$.  Note that in order for the test to have substantial power, the $P^{2}$ must be substantially smaller than $\Delta$.}
 
 \section{Conclusion}

If none of the explanatory variables in a linear regression analysis are statistically significant, can we simply disregard the full model?  How can we formally test whether the proportion of variance attributable to the full set of explanatory variables is too small to be considered meaningful?  In this short article, we introduced a non-inferiority test to help address these questions.  The test can be used to reject effect sizes that are as large or larger than a pre-determined $\Delta$ as estimated by $R^{2}$.  Note that researchers must decide which effect size is considered meaningful or relevant \citep{lakens2018equivalence}, and define $\Delta$ accordingly, prior to observing any data; see \cite{campbell2018make} for details.

The non-inferiority test put forward is specifically intended for the case of random regressors which is a common case in the social sciences and in observational research more broadly.  As such, this paper supplements the work of \cite{campbelllakens2019} who put forward a non-inferiority testing of the coefficient of determination in a linear regression with fixed regressors.  It would be worthwhile to investigate the extent to which the two tests differ.  It would also be worthwhile to expand upon the very limited simulation study from Section 3.  A larger simulation study to further our understanding of how the non-inferiority test operates in a variety of scenarios would be valuable.

\vskip 0.52in

\pagebreak

 \section{Appendix: R-code}
 \noindent Note that one can calculate the confidence interval from equation (\ref{eq:CI}) and the $p$-value from equation (\ref{eq:pval}) in \verb|R| with the following R code.

\noindent An R function for calculating the confidence interval from equation (\ref{eq:CI}):
 
  \begin{verbatim}
UpperCI_random <- function(Rsq, n, k, alpha, tol = 1.0e-12){
    Psq <- Rsq; Psq_last <- 1;	 # initial value
    while(abs(Psq_last - Psq) > tol){
        Psq_last  <- Psq
        v 		  <- (((n-k-1)*Psq + k)^2)/(n-1-(n-k-1)*(1-Psq)^2)
        Fstat 	  <- qf(alpha/2, v, n-k-1)
        Psq_num   <- (n-k-1)*Rsq - (1-Rsq)*k*Fstat
        Psq_den   <- (n-k-1)*(Rsq + (1-Rsq)*Fstat)
        Psq 	  <- Psq_num/Psq_den}
    UpperCI <- ((n-k-1)*Rsq - (1-Rsq)*k*Fstat) / ((n-k-1)*( Rsq + (1-Rsq)*Fstat)) 
return(UpperCI)}

## Example: a 90% upper CI for P2 with N=1250, K=6, R2=0.085:
N <- 1250; K <- 6; Rsquared <- 0.085; Alpha <- 0.10;
UpperCI_random(Rsq = Rsquared, n = N, k = K, alpha = Alpha)
# 0.1069415
# we can compare this to the CI based on the scaled noncentral F approximation:
library("MBESS")
CI_compare <- ci.R2(R2=Rsquared, K, N-K-1, TRUE, conf.level=1-2*Alpha)
CI_compare$Upper.Conf.Limit.R2
# 0.1013726
  \end{verbatim}

\noindent An R function for calculating the $p$-value from equation (\ref{eq:pval}) :
  \begin{verbatim}
noninvR2_random <- function(Rsq, n, k, delta, tol = 1.0e-12){

    Psq      <- Rsq; Psq_last <- 1; # initial value
    F_num 	 <- (n-k-1)*Rsq*(delta-1)
    F_den 	 <- ((Rsq-1) * (delta*(n-k-1) + k))
    Fstat 	 <- F_num/F_den

    while(abs(Psq_last - Psq) > tol){
        Psq_last <- Psq
        v        <- (((n-k-1)*Psq + k)^2)/(n-1-(n-k-1)*(1-Psq)^2)
        Psq_num  <- (n-k-1)*Rsq - (1-Rsq)*k*Fstat
        Psq_den  <- (n-k-1)*(Rsq + (1-Rsq)*Fstat)
        Psq      <- Psq_num/Psq_den
}
pval <- pf(Fstat, v, n-k-1, lower.tail=TRUE)
return(pval)}

## Example: a non-inferiority test for P2 with N=1250, K=6, R2=0.085 and Delta=0.10:
N <- 1250; K <- 6; Rsquared <- 0.075; Delta <- 0.10
noninvR2_random(Rsq = Rsquared, n = N, k = K, delta = Delta)
# 0.02710537
  \end{verbatim}

  \pagebreak
 
 \vspace{10cm}
 
 \pagebreak

\bibliography{truthinscience} 

\end{document}